\documentclass[12pt]{article}%v rabote 30.09.11
\usepackage{graphicx}
\usepackage{cite}
\usepackage{amssymb}
\usepackage{amsmath}
\numberwithin{equation}{section}

\newcommand{\eh}{\hfill}\newlength{\sperr}

\newenvironment{proof}{{\settowidth{\sperr}{\bf\rm
Proof}%
\par\addvspace{0.3cm}\noindent\parbox[t]{1.3\sperr}
{\bf\rm P\eh r\eh o\eh o\eh f.\eh }%
}}{\nopagebreak\mbox{}\hfill
$\Box$\par\addvspace{0.3cm}}

\def\wh{\widehat}
\def\Om{\Omega}
\def\Vol{\mathrm{Vol}}
\def\dist{\mathrm{dist}}

\def\s{\sigma}

\def\ve{\varepsilon}
\def\nn{\nonumber}

\def\BR{{\mathbb R}}

\def\clc{{\mathcal C}}
\def\cld{{\mathcal D}}

\newtheorem{Pa}{Paper}[section]
\newtheorem{Tm}[Pa]{{\bf Theorem}}
\newtheorem{Core}{{\bf Conjecture}}[section]

\newtheorem{Cy}{{\bf Corollary}}[section]
\newtheorem{Rk}{{\bf Remark}}[section]

\newtheorem{Ee}{{\bf Example}}[section]
\newtheorem{Dn}{{\bf Definition}}[section]
\newtheorem{Pn}{{\bf Proposition}}[section]
\title{Inhomogeneous 
Boltzmann equations: 
distance, asymptotics and comparison  of the classical and quantum cases}
\author{Lev Sakhnovich}
\date{}
\begin{document}
\maketitle

\thanks{99 Cove ave., Milford, CT, 06461, USA \\
 E-mail: lsakhnovich@gmail.com}\\

% \textbf{Mathematics Subject Classification (2010):} Primary 35Q20, 82B40;
%Secondary  51K99 \\

\begin{abstract}

The notion of distance between a global
Maxwellian function  and an arbitrary solution $f$
(with the same  total density $\rho$ at the fixed moment $t$) of   Boltzmann equation is introduced.
In this way we essentially generalize the important Kullback-Leibler distance, which was used
before. Namely,  we generalize it for the spatially inhomogeneous case.
An extremal problem to find a solution of the Boltzmann equation, such that $\dist\{M,f\}$ is minimal in the class of solutions with the fixed values
of energy and of $n$ moments, is solved. The cases of the classical and quantum (for Fermi and Bose particles) Boltzmann equations are studied and compared. The asymptotics and stability
of solutions of the  Boltzmann equations are also considered.
\end{abstract}

\section{Introduction.}
We  consider the classical and quantum versions of Boltzmann
 equations (where the quantum version contains both the fermion and boson cases).
The important notion of Kullback-Leibler distance
\cite{KL}, which was fruitfully used
before (see further references in the recent papers \cite{Hab, SH, Vi2}),
 is essentially generalized and new
conventional extremal problems, which appear in this way, are solved. The solution $f(t,x,\zeta)$
of the Boltzmann equation is studied in the bounded domain
$\Omega$ of the $x$-space. Such an approach essentially changes the usual situation, that is,
the total energy depends on $t$ and
the notion of distance between
a stationary solution and an  arbitrary solution of the Boltzmann equation
 includes the $x$-space.  Thus,  the notion of distance remains well-defined in the spatially inhomogeneous case  too. Recall that the Kullback-Leibler distance is  defined only in the spatially homogeneous case.
The comparison of the classical and quantum mechanics, which was treated in \cite{LAS1, LAS2, LAS3}, is generalized here for the case of the Boltzmann equations. It is especially interesting for the applications that the fermion and  boson cases are essentially different from this point of view.
In the last section of the paper we introduce the dissipative and conservative solutions and  find the conditions under which the stationary solution of the classical Boltzmann equation is  stable.

 First, we discuss the classical case. The well-known classical  Boltzmann equation for the monoatomic gas has the form
 \begin{equation}\label{1}
 \frac{\partial{f}}{\partial{t}}=-\zeta{\cdot}\triangledown_{x}f+
 Q(f,f),\end{equation}
 where  $t{\in}\BR$ stands for time,   $x=(x_{1},\ldots ,x_{n}){\in}\Omega$ stands for
 space coordinates,  $\zeta=(\zeta_{1},\ldots ,\zeta_{n}){\in}\BR^{n}$ is velocity,
 and $\BR$ denotes the real axis. The collision
 operator $Q$ is defined by the relation
 \begin{equation}\label{1'}
 Q(f,f)=\int_{\BR^{n}}\int_{S^{n-1}}B(\zeta-\zeta_{\star},\sigma)[f(\zeta^{\prime})f(\zeta^{\prime}_{\star})-
 f(\zeta)f(\zeta_{\star})]{d\sigma}d\zeta_{\star},
 \end{equation}
 where $B(\zeta-\zeta_{\star},\sigma){\geq}0$  is the collision kernel.
 Here we used the notation
 \begin{equation}\label{2}
 \zeta^{\prime}=(\zeta_{\star}+\zeta)/2+\sigma|\zeta_{\star}-\zeta|/2,\,
 \zeta^{\prime}_{\star}=(\zeta_{\star}+\zeta)/2-\sigma|\zeta_{\star}-\zeta|/2,
 \end{equation}where $\sigma{\in}S^{n-1}$, that is, $\sigma{\in}\BR^{n}$
 and $|\s|=1$.
 The solution $f(t,x,\zeta)$ of Boltzmann equation \eqref{1} is the distribution function of gas. We start with some global
Maxwellian function $M$, which is the stationary solution (with the total density $\rho$) of the Boltzmann equation.
The notion of distance between the global
Maxwellian function  and an arbitrary solution $f$
(with the same  value  $\rho$ of the total density at the fixed moment $t$) of the Boltzmann equation is introduced.
As already mentioned before, our approach enables us to treat also the inhomogeneous case.
 An extremal problem to find a solution of the Boltzmann equation, such that
 $\dist\{M,f\}$ is minimal in the class of solutions with the fixed values
of energy and of $n$ moments, is solved.

The same considerations prove fruitful for the quantum Boltzmann equation.
Our definition  of the quantum entropy $S_q$ is slightly different from the previous definitions (see \cite{Dol, Lu2}).
We  show that the natural requirement
\begin{equation}\label{q1}
S_{q}{\to}S_{c},\quad \varepsilon {\to}0 \quad (S_c \,\,\mathrm{is} \,\, \mathrm{the} \,\,
\mathrm{classical} \,\, \mathrm{entropy})
\end{equation}
is not fulfilled in the case of old definition, however \eqref{q1}  holds in the case of   our modified definition (see Section \ref{Se6}).

Some  necessary preliminary definitions and results are given in Section \ref{Prel}.
An important functional, which attains maximum at the global
Maxwellian function is introduced in Section \ref{Extr}.
The distance between solutions of \eqref{1} and the corresponding extremal problem
are studied in Section \ref{Dist}. The modified Boltzmann equations for Fermi and Bose
particles (the quantum cases) are considered in Sections \ref{sec5} and \ref{Se6}.
A comparison of the classical and quantum cases is also conducted in Section \ref{Se6}.
Finally, Section \ref{sec7} is dedicated to the asymptotics and stability of  solutions.

We use the notation $C_{0}^{1}$ to denote the class
of differentiable functions $f(\zeta)$, which tend to zero sufficiently rapidly
when $\zeta$ tends to infinity.
%%%%%%%%%%%%%%%%%%%%%%%%%%%%%%%%%%%%%%%
 \section{Preliminaries: basic definitions and results}\label{Prel}
In this section we  present some  well-known  notions and results
 connected with the Boltzmann equation.
The distribution function $f(t,x,\zeta)$ is non-negative:
\begin{equation}\label{3} f(t,x,\zeta){\geq}0,
\end{equation}
and so the entropy
 \begin{equation}\label{4}
 S(t,f)=-\int_{\Omega}\int_{\BR^{n}}f(t,x,\zeta)\log
  f(t,x,\zeta)d{\zeta}dx
\end{equation}
is well-defined.
\begin{Dn}\label{CInv}
A function $\phi(\zeta)$ is called a collision invariant if it satisfies
 the relation
 \begin{equation}\label{5}
  \int_{\BR^{n}}\phi(\zeta)Q(f,f)(\zeta)d\zeta=0\quad  {\mathrm{for \,\, all}}
  \quad
  f{\in}C_{0}^{1}.
 \end{equation}
\end{Dn}
 It is well-known (see \cite{Vi1}) that there are the following collision invariants:
 \begin{equation}\label{6}
 \phi_{0}(\zeta)=1,\quad
 \phi_{i}(\zeta)=\zeta_{i} \quad(i=1,2,\ldots ,n),\quad
  \phi_{n+1}(\zeta)=|\zeta|^{2}.
  \end{equation}
The notions of density $\rho(t,x)$,  total density $\rho(t)$, mean velocity $u(t,x)$,
energy $E(t,x)$, and total  energy $E(t)$ are introduced via formulas:
 \begin{align}\label{7}&
 \rho(t,x)=\int_{\BR^n} f(t,x,\zeta)d\zeta , \quad \rho(t)=\int_{\Omega}\rho(t,x){dx},\\
 \label{8}&
 u(t,x)=\big(1/\rho(x,t)\big)\int_{\BR^n} \zeta f (t,x,\zeta){d}\zeta,
 \\ \label{9}&
 E(t,x)=\int_{\BR^n} \frac{|\zeta|^{2}}{2}f(t,x,\zeta)d\zeta , \quad
E(t)=\int_{\Omega}\int_{\BR^n} \frac{|\zeta|^{2}}{2}f(t,x,\zeta)d\zeta dx.
\end{align}
The function
\begin{equation}\label{9'}
f=\big(\rho / (2{\pi}T)^{n/2} \big) \exp\big(- |\zeta-u|^{2} /(2T)\big).
\end{equation}
is called the \emph{global Maxwellian} and is a function of the mass density $\rho>0$, bulk velocity $u=(u_{1},\ldots ,u_{n})$ and temperature T.
We assume that the domain $\Omega$ is bounded and so its volume
is bounded too:
\begin{equation}\label{11}
\Vol(\Omega)=V_{\Omega}<\infty.
\end{equation}
Therefore, the function
\begin{equation}\label{9''}
M(\zeta)=\big(\rho / \big(V_{\Omega}(2{\pi}T)^{n/2} \big)\big)
 \exp\big(- |\zeta-u|^{2} /(2T)\big)
\end{equation}
is a global Maxwellian with the constant total density $\rho$.
\begin{Pn}\label{Pn1} \cite{Vi1} The global Maxwellian function $M(\zeta)$ is the stationary solution of the Boltzmann equation \eqref{1}.
\end{Pn}
Boltzmann proved  in \cite{Bol} the  fundamental result below:
\begin{Tm} \label{Tm2} Let $f\in C_{0}^{1}$ be a non-negative solution of equation \eqref{1}. Then the following inequality holds:
\begin{equation}\label{10}
{dS}/{dt}\,\, {\geq} \,\, 0.
\end{equation}
\end{Tm}
%%%%%%%%%%%%%%%%%%%%%%%%%%%%%%%%%%%%%%%%%
\section{Extremal problem}\label{Extr}
Similar to the cases considered in \cite{LAS3, LAS4},
 an important role is played by the functional
\begin{equation}\label{12}
 F(f)=\big(F(f)\big)(t)={\lambda}{E}(t)+S(t),\quad \lambda=-1/T,\end{equation}
 where $S$ and ${E}$, respectively, are defined by formulas \eqref{4} and \eqref{9},
 and the functional \eqref{12} is considered on  the class of functions with the same $\rho(t)=\rho$ at the fixed moment $t$.
 The parameters $\lambda=-1/T$ and $\rho$ are fixed. 
 
 Now,   we use the calculus
of variations (see \cite{Ha}) and find the function $f_{\max}$ which maximizes
the functional \eqref{12}. %under additional condition
%\begin{equation}\label{13}
%\rho=\int_{\Omega}\rho(t,x){dx}.\end{equation}
 The corresponding Euler's equation takes the form
\begin{equation}\label{14}
\frac{\delta}{\delta{f}}\Big[{\lambda}\frac{|\zeta|^{2}}{2}f-f\log f+
{\mu}f\Big]=0.\end{equation}%\label{15}
Here $\frac{\delta}{\delta{f}}$ stands for the functional derivative.
Our extremal problem is conditional and
 $\mu$ is the Lagrange multiplier. Hence, we have
\begin{equation}\label{16}
{\lambda}\frac{|\zeta|^{2}}{2}-1- \log{f}+
{\mu}=0.
\end{equation}
From the last relation we obtain
\begin{equation}\label{17}
f=Ce^{-|\zeta|^{2}/(2T)}.
\end{equation}
Formulas \eqref{9''} and \eqref{17}
imply that
\begin{equation}\label{18}
f=M(\zeta)=\frac{\rho}{V_{\Omega}(2{\pi}T)^{n/2}}e^{-\frac{|\zeta|^{2}}{2T}}.
\end{equation}
In view of \eqref{4}, \eqref{9}, and \eqref{12} we see that
\begin{equation}\label{!}
F(f)=\int_{\Omega}\int_{\BR^{n}}
L_f(t,x,\zeta)d{\zeta}dx,
\quad L_f=-\Big(\frac{|\zeta|^{2}}{2T}+\log f\Big)f.
\end{equation}
For  positive $f$ (including the case $f=M$) and for $L_f$ given in \eqref{!}, we have the inequality
\begin{equation}\label{19}
\frac{\delta^{2}}{\delta{f}^{2}}L_f=-1/f<0.
\end{equation}
\begin{Cy}\label{Cy}
The global Maxwellian function $M(\zeta)$, which is
defined by formula \eqref{17}, gives the maximum of the functional  $F$
on the class of functions with the same value $\rho$
of the total density $\rho(t)$ at the fixed moment $t$.
\end{Cy}
It follows from \eqref{7}, \eqref{18}, and \eqref{!} that
\begin{equation}\label{4.3}
F(M)=-{\rho}\log\Big(\frac{\rho}{V_{\Omega}(2{\pi}T)^{n/2}}\Big).
\end{equation}
Therefore, Corollary \ref{Cy}
can also be  proved without using the calculus of variation  (see \cite{TVi}).
Indeed, taking into account
relations \eqref{18}, \eqref{!}, and \eqref{4.3} and the fact that
 the total densities of $M$ and $f$ are equal, we have
\begin{equation}\label{20}
F(M)-F(f)=\int_{\Omega}\int_{\BR^{n}}
M\Big(1-\frac{f}{M}+\frac{f}{M}\log\frac{f}{M}\Big)d{\zeta}dx.
\end{equation}
Using inequality $1-x+x\log{x}>0$ for $x>0,\,\, x{\ne}1$, we derive from
\eqref{20} that
\begin{equation}\label{2!}
F(M)-F(f)>0
\quad (f{\ne}M).
\end{equation}
\begin{Rk} \label{Rk3.1} Since the extremal problem is conditional, the connection between the energy and entropy can be interpreted in terms of game theory.
The  functional  \eqref{12} defines this game. The global Maxwellian function $M(\zeta)$ is the solution of it.  A game interpretation of quantum and classical mechanics problems is given in the papers \cite{LAS3, LAS4}.
\end{Rk}
\begin{Rk} \label{Rk3.2}
Inequality \eqref{4.3} is valid for all the non-negative functions $f$ with the fixed density $\rho$
at $t$ (not necessarily solutions of the Boltzmann equation).
\end{Rk}
%%%%%%%%%%%%%%%%%%%%%%%%%%%%%%%%%%%%%%

\section{Distance}\label{Dist}
Let $f(t,x,\zeta)$ be a nonnegative solution of the Boltzmann equation \eqref{1}. We assume that $T$ and the value
$\rho=\rho(t)$ at some moment $t$  are fixed. According to \eqref{2!} we have
\begin{equation} \label{4.1}
F(M)-F(f){\geq}0,
\end{equation}
where the global Maxwellian function $M(\zeta)$ is defined in \eqref{18}. The equality in  \eqref{4.1} holds
if and only if $f(t,x,\zeta)=M(\zeta)$. Hence, we can introduce the following
definition of distance between the solution $f(t,x,\zeta)$ and the global Maxwellian function $M(\zeta)$:
\begin{equation}\label{4.2}
\dist\{M,f\}=F(M)-F(f).
\end{equation}
\begin{Rk}\label{Rk4.1}  In the  spatially homogeneous case
(if not only the total densities
$\rho_{M}$ and $\rho_{f}$ of $M$ and $f$  are equal
but the energies $E_{M}$ and $E_{f}$ are equal too),
our definition \eqref{4.2} of distance
coincides with the Kullback-Leibler distance (see \cite{Vi2}).
However, our approach enables us to treat also the inhomogeneous case.
\end{Rk}
Next, we study the case $E_M\ne E_f$ and start with an example.
\begin{Ee} \label{Ee4.2}
Let $T_1\ne T$ and consider the global   Maxwellian function
\begin{equation}\label{4.4}
M_{1}(\zeta)=\frac{\rho}{V_{\Omega}(2{\pi}T_{1})^{n/2}}
\exp\Big(-\frac{|\zeta|^{2}}{2T_{1}}\Big).
\end{equation}
Direct calculation shows that
\begin{align}\label{3!}&
E_1=E_{M_1}=\rho nT_1/2\not=E, \\
\label{4.5}&
F(M_{1})=-{\rho}\left(\log \Big(\frac{\rho}{V_{\Omega}(2{\pi}T_{1})^{n/2}}\Big)-n(1-T_{1}/T)/2\right).
\end{align}
It follows from \eqref{4.3} and \eqref{4.5} that
\begin{equation} \label{4.6}
\dist\{M,M_{1}\}=-{\rho}n\big(\log(T_{1}/T)-T_{1}/T+1\big)/2.
\end{equation}
\end{Ee}
We introduce the class $C(\rho,E_{1},U)$ of non-negative functions
functions $f(t,x,\zeta)$ with the given total density $\rho$ (see (2.8)), total energy
\begin{equation}\label{4.7}
\int_{\Omega}\int_{\BR^n}\frac{|\zeta|^{2}}{2}f(t,x,\zeta)d{\zeta}dx=E_1,
\end{equation}
and total moments $U=\big(U_{1},U_{2},\ldots ,U_{n}\big)$,
where
\begin{equation}\label{4.8}
U_{k}=\int_{\Omega}\int_{\BR^n}{\zeta_{k}}f(t,x,\zeta)d{\zeta}dx.
\end{equation}
Recall that  the global Maxwellian function $M$ is defined by \eqref{18}. \\

\noindent\textbf{Extremal problem.} \emph{Find  a function $f$, which  minimizes
the functional} $\dist\{M,f\}$ \emph{on the class} $C(\rho,E_{1},U)$. \\

The corresponding Euler's equation takes the form
\begin{equation}\label{4.9}
\frac{\delta}{\delta{f}}\Big[({\lambda}+\nu)\frac{|\zeta|^{2}}{2}f-f\log f+
{\mu}f+f\sum_{k}\gamma_{k}\zeta_{k}\Big]=0.\end{equation}
Recall that our extremal problem is conditional, and
 $\mu,\, \nu,\, \gamma_{k}$ are  the Lagrange multipliers. Hence, we have
\begin{equation}\label{4.10}
({\lambda}+\nu)\frac{|\zeta |^{2}}{2}-\log f-1
+{\mu}+\sum_{k}\gamma_{k}\zeta_{k}=0.
\end{equation}
From the last relation we obtain
\begin{equation}\label{4.11}
f=C\exp\Big((\lambda+\nu)\frac{|\zeta|^{2}}{2}+\sum_{k}\gamma_{k}\zeta_{k}\Big).
\end{equation}
According to \eqref{7} we have $\lambda +\nu<0$. Now, we rewrite \eqref{4.11} as
\begin{equation}\label{4!}
f=C_1\Big(-\frac{2\pi}{\lambda + \nu} \Big)^{-n/2}
\exp\Big(\frac{\lambda+\nu}{2}\sum_{k}\Big(\zeta_k+\frac{\gamma_{k}}
{\lambda+\nu}\Big)^2\Big),
\end{equation}
where
\begin{equation}\label{4.13}
C_{1}=C\frac{\pi^{n/2}}{(-(\lambda+\nu)/2)^{n/2}}\exp\Big(-\frac{\sum_{k}\gamma_{k}^{2}}{2(\lambda+\nu)}\Big).
\end{equation}
To calculate the parameters $\mu,\nu,\gamma_{k}$ we use again the well-known formulas
\begin{equation}\label{4.12}
\int_{-\infty}^{\infty}e^{-a\xi ^{2}}d\xi=\sqrt{\pi/a},\quad
\int_{-\infty}^{\infty}\xi^{2}e^{-a\xi^{2}}d\xi=\frac{1}{2a}\sqrt{\pi/a},\quad a>0.
\end{equation}
Formulas \eqref{7}, \eqref{4.7}, \eqref{4.8}, \eqref{4!}, and \eqref{4.12}
imply that
\begin{equation}\label{4.14}
C_{1}=\rho/V_{\Omega},\quad \gamma_{k}/(\lambda+\nu)=-U_{k}/\rho,
\quad -(\lambda+\nu)=T_{1}^{-1},
\end{equation}
where
\begin{equation}\label{4.15}
T_1=\frac{2}{n\rho}E_1-\frac{1}{n\rho^2}\sum_k U_k^2.
\end{equation}
Because of \eqref{4!} and \eqref{4.14} we see that $f$ is just another
global Maxwellian function
\begin{equation}\label{4.17}
f=M_{2}(\zeta)=\frac{\rho}{V_{\Omega}(2{\pi}T_{1})^{n/2}}
\exp\Big(-\frac{|\zeta-U/{\rho}|^{2}}{2T_{1}}\Big).
\end{equation}
In the same way as \eqref{4.5} we obtain:
\begin{equation}\label{4.17'}
F(M_{2})=-{\rho}\left(\log\Big(\frac{\rho}{V_{\Omega}(2{\pi}T_{1})^{n/2}}\Big)-n(1-T_{1}/T)/2\right)-
\frac{1}{2\rho T}|U|^{2}.
\end{equation}
Moreover, formulas \eqref{!} and \eqref{4.2} imply the relations
\begin{equation}\label{4.16}
\dist\{M,f\}=\int_{\Omega}\int_{\BR^{n}}\big(L_M(t,x,\zeta)-
L_f(t,x,\zeta)\big)d{\zeta}dx, \quad
\frac{\delta^{2}}{\delta{f}^{2}}(L_M-L_f\big)=1/f.
\end{equation}
That is, the functional $\dist\{M,f\}$ attains its minimum on the function $f=M_2$, which satisfies conditions $\rho(t)=\rho$, \eqref{4.7}, and \eqref{4.8}. More precisely, in view of \eqref{4.17'}
we have
\begin{equation}\label{4.16'}
\dist\{M,M_2\}=-\frac{n{\rho}}{2}\big(\log(T_{1}/T)-T_{1}/T+1\big)+\frac{|U|^2}{2\rho T}.
\end{equation}
Hence, the following assertion is valid.
\begin{Pn}
\label{Pn4.1} Let $M$ and $M_{2}$, respectively, be defined by \eqref{18} and \eqref{4.17}. If the function $f$ satisfies conditions $\rho(t)=\rho$,
\eqref{4.7}, \eqref{4.8}, and $f{\ne}M_{2}$, then
\begin{equation}\nn
\dist\{M,f\}>-\frac{n{\rho}}{2}\big(\log(T_{1}/T)-T_{1}/T+1\big)+\frac{|U|^2}{2\rho T}.
\end{equation}
\end{Pn}
\begin{Dn} \label{DnMax} We denote by $\widehat{M}$ the Maxwell function of the form \eqref{18}, where $\rho=(1/e)(2\pi T)^{n/2}V_{\Omega}$.
\end{Dn}
According to \eqref{4.3} we have
\begin{equation}F(\widehat{M})>F(M),\quad M{\ne}\widehat{M}.\end{equation}
Hence the following statement is valid.
\begin{Pn}\label{PnMax}  The inequality
\begin{equation}\label{4.20}
\widehat{G}{(f)}=F(\widehat{M})-F(f)>0,\quad f{\ne}\widehat{M}
\end{equation}
is fulfilled  for all non-negative $f$.
\end{Pn}
We call $\widehat{G}$ in \eqref{4.20} {\it the Lyapunov functional}, and will study it
in greater detail in Section \ref{sec7}.
%%%%%%%%%%%%%%%%%%%%%%%%%%%%%%%%%%%%%%%%%%%%%
%%%%%%%%%%%%%%%%%%%%%%%%%%%%%%%%%%%%%%%%%%%%%
%%%%%%%%%%%%%%%% nachalo %%%%%%%%%%%%%%%%%%%

 \section{Modified Boltzmann equations for Fermi and Bose particles. }\label{sec5}

 We study the modified Boltzmann equation which takes into account the
quantum effect  \cite{Dol, Lu1}:
 \begin{equation}\label{5.1}
 \frac{\partial{f}}{\partial{t}}=-\zeta{\cdot}\triangledown_{x}f+
 C (f,f).
 \end{equation}
 The collision
 operator $C$ is defined by the relation
\begin{align}\notag
C(f,f) =\int_{\BR^{n}}\int_{S^{n-1}}& B(\zeta-\zeta_{\star},\sigma)
\big[ f\left(\zeta^{\prime}\right) f\left( \zeta^{\prime}_{\star}\right)\left( 1+{\varepsilon}f(\zeta)\right)\left( 1+{\varepsilon}f\left(\zeta_{\star}\right)\right)
\\  \label{5.2} & -
 f(\zeta)f\left( \zeta_{\star}\right)\left( 1+{\varepsilon}f\left(\zeta^{\prime}\right)\right)\left( 1+{\varepsilon}f\left(\zeta^{\prime}_{\star}\right)\right)  \big]
{d\sigma}d\zeta_{\star},
\end{align}
where $\zeta^{\prime}$ and
$ \zeta^{\prime}_{\star}$ are introduced in \eqref{2}, and  $\varepsilon \in \BR$. If $\ve =0$, the right-hand side of \eqref{5.2} coincides with the right-hand side of \eqref{1'},
that is, we get the classical case.  The inequalities $\varepsilon>0$ and $\varepsilon<0$
hold for bosons and fermions, respectively. Similar to the classical case
the quantum density $\rho_{\ve}$ and quantum energy $E_{\ve}$ are given by formulas \eqref{6} 
and  \eqref{9}, respectively. However, the quantum entropy $S(t,\ve)$ ($\ve \not=0$) is defined in a 
more complicated way:
 \begin{equation} \label{5.3}
 S(t,f,\varepsilon)=-\int_{\Omega}\int_{\BR^{n}}[f\log f-(1/\varepsilon)(1+{\varepsilon}f)
\log{( 1+{\varepsilon}f)}+f]d{\zeta}dx
\end{equation}

\begin{Rk}\label{Rk5.1}
Our definition \eqref{5.3} of entropy is slightly different from the previous definitions (see \cite{Dol, Lu2}). Namely, formula \eqref{5.3} contains the additional summand
 \begin{equation} \label{5.5}
  -\rho_{\ve}=-\int_{\Omega}\int_{\BR^{n}}fd{\zeta}dx.
  \end{equation}
We shall show that the natural requirement
 \begin{equation} \label{5.6}
 S(\ve){\to}S_{c},\quad \varepsilon {\to}0
 \end{equation}
 is fulfilled only in the case that \eqref{5.3} holds.
 \end{Rk}

%%%%%%%%%%%%%%%%%%%%%%%%%%%%%%%%%%%%%%%
%%%%%%%%%%%%%%%%%%%%%%%%%%%%%%%%%%%%%%%%%

\section{Modified extremal problem} \label{Se6}

\paragraph{1.}
We assume again that the domain $\Omega$ is bounded and introduce the functional
\begin{equation} \label{6.1}
F_{\varepsilon}(f)={\lambda}{E_{\varepsilon}}(f)+S(f,\varepsilon),\quad\lambda=-1/T,
\end{equation}
 where  ${E_{\varepsilon}}(f)$ and $S(f,\varepsilon)$ are defined by formulas \eqref{9} and \eqref{5.3} respectively.
The parameters $\lambda=-1/T$ and $\rho$ are fixed.

 Again   we use the calculus
of variations (see \cite{Dol}) and find the function $f_{max}$ which maximizes
the functional \eqref{6.1}
under additional condition
 \begin{equation} \label{6.2}
 \int_{\Omega}\rho(t,x){dx}=\rho.
 \end{equation}
  The corresponding Euler's equation takes the form
 \begin{equation} \label{6.3}
 {\lambda}\frac{|\zeta|^{2}}{2}- \log{f}+\log(1+{\varepsilon}f)-1+
{\mu}=0.
\end{equation}
From the last relation we obtain
 \begin{equation} \label{6.4}
 f/(1+{\varepsilon}f)=Ce^{-\frac{|\zeta|^{2}}{2T}}.
 \end{equation}
 Formula \eqref{6.4} implies  that
 \begin{equation} \label{6.5}  f=M_{\varepsilon}=\frac{Ce^{-\frac{|\zeta|^{2}}{2T}}}
{1-C{\varepsilon}e^{-\frac{|\zeta|^{2}}{2T}}}.
\end{equation}
It is required that the distribution $M_{\ve}$ is positive, that is,
 \begin{equation} \label{6.5'}  
 C>0, \qquad -\infty<C{\varepsilon} \leq 1, 
\end{equation}
and further we assume that \eqref{6.5'} holds. Moreover, \eqref{6.5'} yields also
the positivity of $1+\ve M_{\ve}$:
 \begin{equation} \label{6.5''}  
M_{\ve}(\zeta)>0, \qquad 1+\ve M_{\ve}(\zeta)>0.
\end{equation}
According to \eqref{7} and \eqref{6.2}, the constant $C$ is defined by the equality
 \begin{equation} \label{6.6}
 V_{\Omega}\int_{\BR^{n}}\frac{Ce^{-\frac{|\zeta|^{2}}{2T}}}
{1-{\varepsilon}Ce^{-\frac{|\zeta|^{2}}{2T}}}d{\zeta}=\rho.
\end{equation}

In view of  \eqref{6.1}, we have the relation which is similar to \eqref{!}:
\begin{equation}\label{q!}
F_{\ve}(f)=\int_{\Omega}\int_{\BR^{n}}
L_{f,\ve}(t,x,\zeta)d{\zeta}dx.
\end{equation}
Though the function $L_{f,\ve}$ is more complicated than $L_f$ in \eqref{!},
we easily get an analog of \eqref{19}:
 \begin{equation} \label{6.7}  \frac{\delta^{2}}{\delta{f}^{2}}L_{f,\ve}
 =-\frac{1}{f(1+{\varepsilon}f)}<0,
 \end{equation}
 which clearly holds if  $f$ and $1+\ve f$ are positive, including the case that $f=M_{\ve}$.
 \begin{Cy}\label{qEPm}
The functional $F_{\ve}$ given by \eqref{6.1} attains its maximum 
$($for positive functions $f$ satisfying  condition \eqref{6.6}$)$ on the function $M_{\ve}$  
of the form \eqref{6.5}. That is, for the distance $G_{\ve}$ we get
 \begin{equation} \label{6.8}
 G_{\varepsilon}(f):=F_{\varepsilon}(M_{\varepsilon})-F_{\varepsilon}(f)>0 \quad (f{\ne}
 M_{\varepsilon}).
\end{equation}
\end{Cy}
\begin{Rk}
%\textbf{Remark 6.1.} 
 The global Maxwellians $M_{\ve} $ play an essential role in boson and fermion theories.
When the standard approach is used, they appear in a more complicated way  $($see \cite[Ch.V, sections 52, 53]{6new} and \cite[Ch.1, sections 9, 10]{3new}$)$.
\end{Rk}

\paragraph{2.} Using the spherical coordinates, we 
calculate the integral on the left-hand  side of \eqref{6.6}  
 \begin{equation} \label{6.9}
 \int_{\BR^{n}}\frac{Ce^{-\frac{|\zeta|^{2}}{2T}}}
{1-{\varepsilon}Ce^{-\frac{|\zeta|^{2}}{2T}}}d{\zeta}=\omega_{n-1}C\int_{0}^{\infty}
\frac{r^{n-1}e^{-\frac{r^{2}}{2T}}}
{1-{\varepsilon}Ce^{-\frac{r^{2}}{2T}}}dr, \quad  \omega_{n-1}=\frac{2{\pi}^{n/2}}{\Gamma(n/2)},
\end{equation}
where $\omega_{n-1}$ is the surface  area of the $(n-1)$-sphere of radius $1$,
and  $\Gamma(z)$ is the Euler's Gamma function.
Taking into account \eqref{6.6} and  \eqref{6.9} we obtain
 \begin{equation} \label{6.12}
 (2{\pi}T)^{n/2}V_{\Omega}CL_{n/2}(C\varepsilon)=\rho,
%(2{\pi}T)^{n/2}V_{\Omega}C_{0}=\rho ,
\end{equation}
where
 \begin{equation} \label{6.13}
  L_{n/2}(z)=\frac{2}{(2T)^{n/2}\Gamma(n/2)}\int_{0}^{\infty}\frac{r^{n-1}e^{-\frac{r^{2}}{2T}}}
{1-ze^{-\frac{r^{2}}{2T}}}dr.
\end{equation}
Because of the equality
 \begin{equation} \label{6.11}
 \int_{0}^{\infty}e^{-ar^{2}}r^{n-1}dr=\frac{1}{2}a^{-n/2}\Gamma(n/2)
 \end{equation}
the function  $L_{n/2}(z)$ admits the expansion
 \begin{equation} \label{6.13'}
  L_{n/2}(z)=\sum_{m=1}^{\infty}\big(z^{m-1}/m^{n/2}\big),
\end{equation}
which yields the next  statement.

\begin{Pn}\label{Proposition 6.1.}
The function $L_{n/2}(z)$ monotonically increases for \\ $0<z<1$ and
 \begin{equation} \label{6.14}
 L_{1/2}(1)=L_{1}(1)=\infty, \quad L_{n/2}(1)<\infty \quad \mathrm{for} \quad n>2.
 \end{equation}
 \end{Pn}
\begin{Rk}
It is easy to see that $L_{n/2}(z)=\infty$ for  $z>1$.
\end{Rk}

In view of Proposition \ref{Proposition 6.1.} we have:

\begin{Cy}
%\textbf{Corollary 6.1.}
If $\varepsilon>0$ $($boson case$)$ and either $n=1$ or $n=2$, then
equation \eqref{6.12} has one and only one solution $C$ such that $C>0$, $C\varepsilon<1$,
and so \eqref{6.5'} holds.
\end{Cy}

\begin{Cy}
%\textbf{Corollary 6.2.}
If $\varepsilon>0$ $($boson case$)$, $n>2$ and
 \begin{equation} \label{6.15}
 (2{\pi}T)^{n/2}V_{\Omega}L_{n/2}(1)>\varepsilon\rho,
 \end{equation}
{then
equation \eqref{6.12} has one and only one solution $C$ such that $C>0$ and $C\varepsilon<1$.}
If, instead of \eqref{6.15}, we have  $(2{\pi}T)^{n/2}V_{\Omega}L_{n/2}(1)=\varepsilon\rho$,
then the solution of \eqref{6.12} is given by $C=1/\ve$ and the corresponding $M_{\ve}$
has singularity at $\zeta =0$.

\end{Cy}

\begin{Rk}
%\textbf{Remark 6.2.}
The function $L_{n/2}(z)$ belongs to the class of the $L$-functions \cite{LaGa} and
is connected with the famous Riemann zeta-function 
 \begin{equation} \label{6.16}
 \zeta(z)=\sum_{k=1}^{\infty}\frac{1}{k^{z}};\quad \Re{z}>1
 \end{equation}
 by the relation
 \begin{equation} \label{6.17}
 L_{n/2}(1)=\zeta(n/2).
 \end{equation}
Hence, some useful estimates for $L_{n/2}(1)$ follow. In particular, we get
 \begin{align} \label{Lev}
 L_{3/2}(1)=2.612,\quad L_{2}(1)=1.645,
 \quad
 L_{5/2}(1)=1.341,\quad L_{3}(1)=1.202.
\end{align}
\end{Rk}

Let us consider the fermion case (i.e., the case $\varepsilon<0$). The next proposition
easily follows from \eqref{6.13} and monotonical increase of $ax(1+ax)^{-1}$ ($a>0$)
on the positive half-axis.

\begin{Pn}\label{Prop6.2} Let $\varepsilon<0$. Then
the function $CL_{n/2}(C\varepsilon)$ monotonically increases with   respect to $C>0$.
Furthermore,  we have $CL_{n/2}(C\varepsilon){\to}\infty$ for $C{\to}\infty$. 
\end{Pn}

\begin{Cy}
%\textbf{Corollary 6.3.}
{If $\varepsilon<0$ $($fermion case$)$, then
equation \eqref{6.12} has one and only one solution $C$ such that} $C>0.$
\end{Cy}

\paragraph{3.}
Consider now
 the   energy for the global Maxwellian $M_{\ve}$:
 \begin{equation} \label{6.20}
 E_{\ve}(M_{\ve})=\int_{\Omega}\int_{\BR^{n}}\frac{|\zeta|^{2}Ce^{-\frac{|\zeta|^{2}}{2T}}}
{1-{\varepsilon}Ce^{-\frac{|\zeta|^{2}}{2T}}}d{\zeta}dx/2=V_{\Omega}\omega_{n-1}C\int_{0}^{\infty}
\frac{r^{n+1}e^{-\frac{r^{2}}{2T}}}
{1-{\varepsilon}Ce^{-\frac{r^{2}}{2T}}}dr/2.
\end{equation}
Formulas \eqref{6.9}--\eqref{6.13} and \eqref{6.20}
 imply that
 \begin{equation} \label{6.21}
 E_{\ve}(M_{\ve})=\left(\frac{n{\rho}T}{2}\right)\frac{L_{n/2+1}(C\varepsilon)}{L_{n/2}(C\varepsilon)}.
\end{equation}
According to \eqref{3!} the corresponding classical energy $E=E_0=E_c$ is given by the formula
 \begin{equation} \label{6.22}
 E_{c}(M)=\frac{n{\rho}T}{2} \qquad (M=M_0).
 \end{equation}

 \begin{Pn}\label{Proposition 6.3.}
 If $\ve >0$ $($boson case$)$, then we have   
 \begin{equation} \label{6.24}
 E_{\ve}<E_{c}.
 \end{equation}
If $\ve<0$ $($fermion case$)$ and
 \begin{equation} \label{6.23'}
  \mathrm{either} \quad n\geq 2, \quad 
 -C\varepsilon{\leq}1 \quad \mathrm{or} \quad
 n=1, \quad -C\ve< 3^{3/2}/2^{5/2}\approx 0.91,
 \end{equation}
 then we have
 \begin{equation} \label{6.24'}
 E_{c}<E_{\ve}.
 \end{equation}
  \end{Pn}
\begin{proof}
%\emph{Proof.}
Taking into account \eqref{6.13'},  we obtain $L_{n/2+1}(C\varepsilon)/L_{n/2}(C\varepsilon)<1$
for $\ve>0$. Hence, in view of \eqref{6.21} and \eqref{6.22}
the inequality  \eqref{6.24}  holds in the boson case.

If $\ve<0$ and conditions \eqref{6.23'} hold, the inequalities
$$L_{n/2}(C\varepsilon)>0 \quad\mathrm{and} \quad L_{n/2+1}(C\varepsilon)-L_{n/2}(C\varepsilon)>0$$ 
follow  from  \eqref{6.13} and \eqref{6.13'}, respectively, and we get 
$$L_{n/2+1}(C\varepsilon)/L_{n/2}(C\varepsilon) \nobreak >1.$$
That is, in view of \eqref{6.21} and \eqref{6.22} the inequality  \eqref{6.24'}  is proved in the fermion
case.
\end{proof}

\paragraph{4.}
For the classical case $\ve=0$ formula \eqref{6.12} (see also \eqref{9''}) implies
 \begin{equation} \label{s5}
 C=C_0=\rho/V_{\Om}\big(2\pi T\big)^{n/2}.
 \end{equation}
In view of \eqref{12}, \eqref{4.3}, and \eqref{s5} we easily derive for $M=M_0$ that 
 \begin{equation} \label{6.26}
 S_{c}=\frac{1}{T}E_{c}-{\rho}\log{C_{0}}.
 \end{equation}
 To calculate the quantum entropy $S(M_{\ve}, \ve)$ we recall \eqref{6.5}
 and use equalities
  \begin{equation} \label{s6}
M_{\ve}=g/(1-\ve g), \quad 1+\ve M_{\ve}=(1-\ve g)^{-1}, \quad g:=Ce^{-|\zeta|^2/(2T)}
\end{equation}
to simplify the expression, which stands  under integral on the right-hand side
of \eqref{5.3} and which we denote by $L_S$:
  \begin{equation} \label{s7}
L_S=M_{\ve}(1+\log g)+(1/\ve)\log(1-\ve g).
\end{equation}
Substitute $\log g=\log C -(1/2T)|\zeta|^2$ into \eqref{s7} and substitute \eqref{s7} into
\eqref{5.3} to get
 \begin{equation} \label{s8}
 S(M_{\ve}, \ve)=\frac{1}{T}E_{\ve}-(1+\log{C})\rho -\frac{1}{\ve}V_{\Om}\int_{\BR^n}
 \log(1-\ve g)d\zeta.
\end{equation}
Using integration by parts and the definition \eqref{9} of energy we rewrite \eqref{s8}:
 \begin{equation} \label{6.27}
 S(M_{\ve}, \ve)=\frac{1}{T}E_{\ve}-(1+\log{C})\rho
+\frac{2E_{\ve}}{nT}.
\end{equation}
From \eqref{6.1}, \eqref{6.22}, \eqref{6.26}, and \eqref{6.27} we see that
 \begin{align}& \label{6.28}
 S(M_{\ve}, \ve)-S_{c}=\frac{n+2}{nT}(E_{\ve}-E_{c})-{\rho}\log({C}/{C_{0}}),
\\
& \label{6.29}  F_{\ve}-F_{c}=\frac{2}{nT}(E_{\ve}-E_{c})-{\rho}\log({C}/{C_{0}})
 \qquad (F_c=F_0).
\end{align}
The behavior of $C$ is of interest and we start with the proposition below.
\begin{Pn} The following inequalities are valid:
 \begin{equation} \label{6.30}
 C>C_{0}\quad \mathrm{for} \,\, \varepsilon<0; \qquad C<C_{0} \quad \mathrm{for} \,\, \varepsilon>0.
 \end{equation}
\end{Pn}
\begin{proof}
According  to \eqref{6.13} and \eqref{6.13'} we have
\begin{align}& \label{s9}
L_{n/2}(z_1)<L_{n/2}(0)=1<L_{n/2}(z_2) \quad \mathrm{for} \quad z_1<0<z_2\leq 1.
\end{align}
Therefore, it is immediate that
\begin{align}& \label{s10}
C_0L_{n/2}(C_0\ve_1)<C_0, \quad C_0<C_0 L_{n/2}(C_0 \ve_2) \quad \mathrm{for} \quad \ve_1<0<\ve_2.
\end{align}
In view of Propositions \ref{Proposition 6.1.} and \ref{Prop6.2} the functions 
$C L_{n/2}(C \ve_1)$ and $C L_{n/2}(C \ve_2)$ increase with respect to $C>0$, and so formulas \eqref{6.12}
and \eqref{s10} imply \eqref{6.30}.
 \end{proof}
 It is immediate from \eqref{6.30} that $C$ is bounded for $\ve>0$. However, $C$ is bounded
 also for the small values of $|\ve|$, when $\ve$ is negative. Indeed, let $-(2C_0)^{-1}<\ve<0$.
Then, formula \eqref{6.13} yields $$2L_{n/2}(2C_0 \ve)>2L_{n/2}(-1)>L_{n/2}(0)=1.$$
Therefore, we have $2C_0L_{n/2}(2C_0 \ve)>C_0$, which in view of 
Proposition \ref{Prop6.2} implies $C<2C_0$.

Now, rewrite \eqref{6.12} as $z=C_0\ve / L_{n/2}(z)$, where $z=C\ve$, and
note that $\left|\frac{d}{dz}\frac{C_0\ve}{L_{n/2}(z)}\right|<1$ for $|z|<1$ and small values of $\ve$. (Since $C$ is bounded,
we see that $|z|<1/2$ for the sufficiently small values of $\ve$.) Thus, we apply iteration method
to the equation $z=C_0\ve /L_{n/2}(z)$ and derive
\begin{align}& \label{s13} 
C=C_0+O(\ve), \qquad \ve \to 0.
\end{align} 
 Next we note that formula \eqref{6.12} yields
 $CL_{n/2}(C\varepsilon)=C_{0}$. Therefore, taking into account \eqref{s13} we get
\begin{align}& \label{s11} 
C/C_{0}=1/L_{n/2}(C\varepsilon)=1-(C_{0}\varepsilon)/2^{n/2}+O(\varepsilon^{2}).
\end{align}
Moreover, from \eqref{s11} we see that 
\begin{align}& \label{s12} 
\log(C/C_{0})=-(C_{0}\varepsilon)/2^{n/2}+O(\varepsilon^{2}).
 \end{align}

Using  relations \eqref{6.13'}, \eqref{6.21}, \eqref{6.22}, and \eqref{s13},
we derive
 \begin{equation} \label{6.31}
 E_{\ve}-E_{c}=-\frac{n{\rho}TC_{0}{\varepsilon}}{{4}(2^{n/2})}+O(\varepsilon^{2}), \quad \ve \to 0.
\end{equation}
Because of \eqref{6.28}, \eqref{6.29}, \eqref{s12}, and \eqref{6.31}, we get the next proposition.
\begin{Pn}
%\textbf{Proposition 6.4.}
For $\ve \to 0$, we have equality \eqref{6.31} as well as equalities below:  
 \begin{align} \label{6.32}&
 S(M_{\ve},\ve)-S_{c}=-\frac{(n-2)\rho C_{0}\varepsilon}{4(2^{n/2})}+O(\varepsilon^{2}), \\
 \label{6.33} &
 F_{\ve}-F_{c}=\frac{{\rho}C_{0}{\varepsilon}}{{2}(2^{n/2})}+O(\varepsilon^{2}).
\end{align}
\end{Pn}

\begin{Cy}
%\textbf{Corollary 6.4.}
Let $\varepsilon_1<0<\ve_2$ be small. Then
 \begin{equation} \label{6.34}
  S(M_{\ve_2},\ve_2)<S_{c}< S(M_{\ve_1},\ve_1) \quad \mathrm{for} \quad n>2,
 \quad F_{\ve_1}<F_{c}<F_{\ve_2} \quad \mathrm{for}\,  \mathrm{all} \quad n.
 \end{equation}
 \end{Cy}
\begin{Rk} We recall that in view of Proposition \ref{Proposition 6.3.} the inequalities
\begin{equation}E_{\ve_2}<E_{c}<E_{\ve_1}, \quad \varepsilon_1<0<\ve_2 \end{equation}
hold without the demand for $\varepsilon_i$ to be small. Here $E_{\ve_2}$ corresponds to the boson
and $E_{\ve_1}$ to the fermion case.
\end{Rk}
\begin{Rk} We note that relations \eqref{6.33} as well as their  physical
interpretation are contained in the well-known book  by
L. Landau and E. Lifshitz \cite[Section 55]{6new}.
\end{Rk}
\begin{Core} Relation \eqref{6.24'}, which was proved for all $-C\varepsilon{\leq}1$ $(\ve<0)$
in the case that  $n\geq 2$, is valid also for all $-C\varepsilon{\leq}1$ $(\ve<0)$
in the case that  $n=1$. We recall that \eqref{6.24'} holds for $n=1$ and $-C\ve< 3^{3/2}/2^{5/2}$.
Moreover, \eqref{6.24'} holds in the extremal case $C\varepsilon=-1$.
Indeed, using \eqref{6.17}, \eqref{Lev}, the relation $\zeta(1/2){\approx}-1.46$ and the well-known equality (see, e.g., \cite[p.17]{LaGa})
\begin{equation} L_{s}(-1)=\zeta(s)(1-2^{1-s}),\end{equation}  
we obtain
\begin{equation}L_{3/2}(-1){\approx}0.765,\quad L_{1/2}(-1){\approx}0.6.\end{equation}
Hence, $L_{3/2}(-1)/L_{1/2}(-1)>1$ and the conjecture is proved for the case that $C\varepsilon=-1$.
\end{Core}
 \section{Lyapunov functional }\label{sec7}
\subsection{Classical case}
In this subsection we prolong the study of the classical Boltzmann equation \eqref{1} and
assume that  $f(t,x,\zeta)$ is its non-negative solution.
Using Gauss-Ostrogradsky formula we write
 \begin{align}& \label{7.1}
 \int_{\Omega}\int_{\BR^{n}}(|\zeta|^{2}/2)\zeta{\cdot}{\bigtriangledown_{x}}fd{\zeta}dx= \int_{\partial\Omega}\int_{\BR^{n}}(|\zeta|^{2}/2)[\zeta{\cdot}n(x)]fd{\zeta}d\sigma=A(t,\Omega),
 \\&
  \label{7.2}  \int_{\Omega}\int_{\BR^{n}}\zeta{\cdot}{\bigtriangledown_{x}}fd{\zeta}dx= \int_{\partial\Omega}\int_{\BR^{n}}[\zeta{\cdot}n(x)]fd{\zeta}d\sigma=B(t,\Omega),
 \end{align}
where $\partial\Omega$ is the piecewise smooth boundary of  $\Omega$, and the integral $\int_{\partial\Omega}g d\sigma$ is the surface integral with $n(x)$ being the outward unit normal to that surface, $x{\in}\partial\Omega.$
\begin{Rk}\label{Rk7.1}
Here $A(t,\Omega)$ and  $B(t,\Omega)$ are
the total energy flux and the total density flux through the surface $\partial\Omega$ per unit time,
respectively. 
\end{Rk}

\begin{Dn}\label{Def7.1}
%\textbf{Definition 7.1.}
We  say that a non-negative solution $f(t,x,\zeta)$
of \eqref{1} belongs  to the class  $\cld(\Omega)$
of  dissipative functions, if
$A(t,\Omega){\geq}0$ for all $t$.
\end{Dn}
\begin{Dn}
%\textbf{Definition 7.2.}
We say that a non-negative solution $f(t,x,\zeta)$
of \eqref{1} belongs  to the class  $\clc(\Omega)$
of conservative functions, if
$A(t,\Omega)=0$ for all $t$.
\end{Dn}
 Clearly we have $\clc(\Omega){\subset}\cld(\Omega).$ We note that the same definitions
 are applicable in the quantum case.
\begin{Pn}
%\textbf{Proposition 7.2.}
If the inequality $f(t,x,\zeta)\geq 0$ and condition
$ f(t,x,\zeta)=f(t,x,-\zeta)$ for $x{\in}\partial\Omega$ hold,
 then we have $f(t,x,\zeta)\in \clc(\Omega)$.
 \end{Pn}
\begin{proof}
Since
$ \int_{\BR^{n}}(|\zeta|^{2}/2)f(t,x,\zeta){\zeta}d{\zeta}=0$, it follows that $A(t,\Om)\equiv 0$ for $A$ 
which is given by \eqref{7.1}.
\end{proof}
\begin{Rk} The so called \textit{bounce-back condition} $ f(t,x,\zeta)=f(t,x,-\zeta)$ means that particles arriving with a certain velocity to the boundary $\partial\Omega$ will bounce back with the opposite velocity (see \cite[p.16]{Vi1}).
\end{Rk}
\begin{Cy}
%\textbf{Corollary 7.1.}
The global Maxwell functions $M$  of the form
\eqref{18}  belong to the conservative class $\clc(\Omega)$.
\end{Cy}
Furthermore, the next  assertion can be easily derived via
direct calculation.
\begin{Cy}
%\textbf{Corollary 7.2.}
The global  Maxwell functions $M$ of the form
\eqref{9''}, also belong to  $\clc(\Omega)$.
\end{Cy}

\begin{Ee}\label{Example 7.1.}
The  well-known and important Maxwellian diffusion example $($see \cite[p.16]{Vi1}$)$
is described by the property
\begin{equation} \label{7.5}
 f(t,x,\zeta)=\rho_{-}(x)M_{b}(\zeta)\quad \mathrm{for} \quad  x{\in}\partial\Omega,
\quad \zeta{\cdot}n(x)>0,
\end{equation}
where $M_{b}(\zeta)$ has the form \eqref{18}.
When we have
 \begin{equation} \label{7.6}
 \int_{\partial\Omega}\int_{\zeta{\cdot}n(x)>0}(|\zeta|^{2}/2)[\zeta{\cdot}n(x)]fd{\zeta}d\sigma {\geq}\int_{\partial\Omega}\int_{\zeta{\cdot}n(x)<0}(|\zeta|^{2}/2)|\zeta{\cdot}n(x)|fd{\zeta}d\sigma,
\end{equation}
the function $f$ in \eqref{7.5} is dissipative. If in relation \eqref{7.6} we have the equality, then $f$  is conservative. Hence, such functions satisfy our statements below $($and the results below are
new even for this case$)$.
\end{Ee}

Now, consider  the Lyapunov functional $\widehat{G}(f)=F(\widehat{M})-F(f)$
for the  equation \eqref{1}.
According to \eqref{4.20} we have
 \begin{equation} \label{7.8}  \widehat{G}{(f)}>0 \quad\mathrm{for} \quad f{\ne}\widehat{M},\quad \widehat{G}(\widehat{M})=0.
 \end{equation}
 Using Theorem \ref{Tm2} we derive the following assertion.
\begin{Tm}\label{Theorem 7.1.}
%\textbf{Theorem 7.1.}
Let $f{\in}C_{0}^{1}$ be a non-negative dissipative solution of  \eqref{1}.  Then the inequality
 $( d\widehat{G}/{dt}){\leq}0$ is valid.
\end{Tm}
\begin{proof}
The function $\phi(\zeta)=|\zeta|^{2}$ is a collision invariant
(i.e., \eqref{5} holds). Therefore, taking into account \eqref{1}, \eqref{7.1}, 
and Definition \ref{Def7.1}  we have
 \begin{equation} \label{7.10}
 \frac{d}{dt}\int_{\Omega}\int_{\BR^{n}}(|\zeta|^{2}/2)fd{\zeta}dx =-A(t,\Omega){\leq}0,
 \end{equation}
 that is, $(dE/dt) \leq 0$. Recall also that $\wh M$ is a stationary solution, and so
$d\widehat{G}/{dt}=-dF(f)/{dt}$.
 Now, the assertion of the theorem follows  from  \eqref{10} and \eqref{12}.
 \end{proof}

% \begin{Cy}
%\textbf{Corollary 7.3.}
According to Theorem \ref{Theorem 7.1.}, if its  conditions are fulfilled and 
$\big(\wh G(f)\big)(t_0)<\delta$, then the inequality $\big(\wh G(f)\big)(t)<\delta$
holds for all $ t>t_{0}$. Thus, the following important result is proved.
\begin{Tm}\label{Imp}
If  the distance is defined by $\widehat{G}$ and $f$ is dissipative,
the stationary solution $\widehat{M}$ is locally stable.
\end{Tm}
 The previous results on local stability \cite{Lu1, TVi} were obtained  for   the spatially homogeneous Boltzmann equation.

\begin{Cy}
%\textbf{Corollary 7.4.}
Let conditions of Theorem \ref{Theorem 7.1.} be fulfilled. Then the function $F(f)$ monotonically increases with respect to $t$ and is bounded. Hence, there is a limit
 \begin{equation} \label{7.13}
 \lim_{t{\to}\infty}F(f)=\Phi{\leq}F(\widehat{M}).
 \end{equation}
 \end{Cy}
Next, assume that  the following limits  exist:
 \begin{align} \label{7.14}
 & \rho_{\infty}=\lim_{t{\to}\infty}\rho(t)\not=0, \quad
U_{\infty}=\lim_{t{\to}\infty}U(t), 
\end{align}
where $\rho(t)$ and $U(t)$ are given by \eqref{7} and  \eqref{4.8}, respectively.
We see from \eqref{4.3} and \eqref{7.14} that the functions $M$ and $M(t)$ of the form \eqref{18}, where
$\rho_{\infty}$ and $\rho(t)$, respectively, are substituted in place of $\rho$, satisfy relations
\begin{align}\label{M1}&
 F(M)=-\rho_{\infty} \log\Big(\frac{\rho_{\infty}}{V_{\Omega}(2{\pi}T)^{n/2}}\Big)= \lim_{t \to \infty} F(M(t)).
\end{align}
\begin{Pn} \label{tosol}
Let the relations  \eqref{7.13} and \eqref{7.14} hold. Then we have the inequality 
 \begin{align} \label{T1}&
F(M)-\Phi \geq |U_{\infty}|^2/(2\rho_{\infty }T).
\end{align}
Moreover, if the inequality \eqref{T1} turns into equality, 
there is a unique Maxwell function $M_{U}$ of the form \eqref{4.17} $($with $\rho = \rho_{\infty}$ and $U=U_{\infty})$
such that
 \begin{equation} \label{7.15'}
 F(M_{U})=\Phi .
 \end{equation}
If the inequality \eqref{T1} is strict, that is,
 \begin{align} \label{T2}&
F(M)-\Phi > |U_{\infty}|^2/(2\rho_{\infty}T),
\end{align}
there are two such functions $(M_1$ and $M_2)$ satisfying
 \begin{equation} \label{7.15}
 F(M_{k})=\Phi \quad (k=1,2).
 \end{equation}
\end{Pn}
\begin{proof} It is immediate that
 \begin{align} \label{T3}&
y(x):=x-1-\log x=0 \quad \mathrm{for} \quad x=1, \quad y(x)>0  \quad \mathrm{for} \quad x>0, \,\, x\not=1.
\end{align}
Since $y\geq 0$, according to Proposition \ref{Pn4.1}  we have
 \begin{align} \label{T4}&
F(M(t))-F(f(t)) \geq |U(t)|/(2\rho(t)T).
\end{align}
In view of \eqref{7.13}-\eqref{M1} and \eqref{T4} we get \eqref{T1}.

Now, using \eqref{4.16'} we rewrite equation \eqref{7.15'} (or, correspondingly, \eqref{7.15})
in the form
 \begin{align} \label{T5}&
\frac{2}{n\rho_{\infty}}\left(F(M)-\Phi-\frac{|U_{\infty}|^2}{2\rho_{\infty} T}\right)=x-1-\log x,
\end{align}
where $M_U$ or, correspondingly, $M_k$ are expressed via solutions $x_k$ of \eqref{T5}
in the form
(compare with \eqref{4.17}):
$$M_U=M_k= \frac{\rho_{\infty}}{V_{\Omega}(2{\pi}x_kT)^{n/2}}
\exp\Big(-\frac{|\zeta-U_{\infty}/{\rho}_{\infty}|^{2}}{2x_kT}\Big).$$
According to \eqref{T3}, the equation \eqref{T5} has a unique solution
when \eqref{T1} turns into equality and has two solutions when \eqref{T1} 
is a strict inequality.
\end{proof}

\begin{Cy}\label{Cy7n1}
%\textbf{Corollary 7.5.}
Let the conditions of Proposition \ref{tosol}  be fulfilled. 
Then
 \begin{equation} \label{7.16}
 F(M_{k})-F(f){\to}0,\quad t{\to}\infty.
 \end{equation}
 \end{Cy}

\begin{Cy}\label{Cy7n2} Let the conditions of Proposition \ref{tosol} 
be fulfilled, where the  strict inequality  \eqref{T2} holds. If the  limit
 \begin{equation} \label{7.14'}
E_{\infty}= \lim E(t)\quad (t{\to}\infty)
 \end{equation}
exists and the corresponding solution 
$f(t,x,\zeta)$ converges to a Maxwell function,
then either $E_{\infty}=E_{1}$ or $E_{\infty}=E_{2}.$
\end{Cy}
\begin{Rk}
\label{Rk7.3} Proposition \ref{tosol}  and Corollaries \ref{Cy7n1} and \ref{Cy7n2} are valid if the limit \eqref{7.13} exists. We do not suppose there, that the corresponding solution $f$ is dissipative.
\end{Rk}
\subsection{Quantum case}
Now, we consider the quantum version \eqref{5.1} of the Boltzmann equation. The corresponding Lyapunov functional $G_{\varepsilon}(f)$                                                                                                                                                                                            has the form \eqref{6.8}.
\begin{Tm}  $($see \cite{Dol} and \cite{TVi}$)$ Let $f$ $(f{\in}C_{0}^{1})$ be a non-negative solution of equation \eqref{6.1}.  Then the inequality
\begin{equation}\frac{dS_{\varepsilon}}{dt}{\geq}0\end{equation}
is valid.
\end{Tm}
In the same way as in the classical case we obtain the assertions.
\begin{Tm}\label{Tm7.3} Let $f\in C_{0}^{1}$ be a non-negative dissipative solution of equation \eqref{6.1}. Then the inequality
\begin{equation}\frac{dG_{\varepsilon}}{dt}{\leq}0\end{equation}
is valid.
\end{Tm}
\begin{Cy} Let the conditions of  Theorem \ref{Tm7.3} be fulfilled. If
\begin{equation}G_{\varepsilon}(f(t_{0},x,\xi))<\delta,\end{equation}
then
\begin{equation}G_{\varepsilon}(f(t,x,\xi))<\delta,\quad t>t_{0}.\end{equation}
\end{Cy}
Thus, we proved that the stationary solution $M_{\varepsilon}$ is \emph{locally stable}, when the distance is defined by $G_{\varepsilon}(f)$ and the function $f$ is dissipative.

\section{Conclusion}
We see that the study of the Boltzmann equations in a bounded domain $\Omega$
and the suggested new extremal problem allow us to introduce a notion of distance
and obtain various results for the inhomogeneous   classical and quantum
cases. In particular, the notion of  the dissipative solutions is introduced and
asymptotics and stability of solutions of the classical and quantum Boltzmann equations 
is studied. Following, e.g., \cite{PlaPla, LAnonext}  we plan also to
consider  solutions of the Boltzmann equations for the case of Tsallis entropy.
The approach could be applied to other related equations, such as the Fokker-Planck
equation.

\end{document}